\documentclass{ieeeaccess}
\usepackage{cite}
\usepackage{amsmath,amssymb,amsfonts}
\usepackage{algorithm, algorithmic}
\usepackage{graphicx}
\usepackage{textcomp}
\usepackage{caption}
\usepackage{qcircuit}
\usepackage{float}

\def\BibTeX{{\rm B\kern-.05em{\sc i\kern-.025em b}\kern-.08em
    T\kern-.1667em\lower.7ex\hbox{E}\kern-.125emX}}

\begin{document}
\history{Date of publication xxxx 00, 0000, date of current version xxxx 00, 0000.}
\doi{10.1109/TQE.2020.DOI}

\vol{draft}
\year{13-9-2024}

\newcommand\numberthis{\addtocounter{equation}{1}\tag{\theequation}}
\newcommand{\Var}[1]{\text{Var}\left(#1\right)}

\title{Estimating shots and variance on noisy quantum circuits}
\author{
  \uppercase{Manav Seksaria}\authorrefmark{1, 2},
  \uppercase{Anil Prabhakar\authorrefmark{1, 2}}
  \address[1]{Centre for Quantum Information, Communication and Computing, Indian Institute of Technology Madras, Chennai, India}
  \address[2]{Department of Electrical Engineering, Indian Institute of Technology Madras, Chennai, India}
}

\begin{abstract}
We present a method for estimating the number of shots required to achieve a desired variance in the results of a quantum circuit. First, we establish a baseline for single-qubit characterisation of individual noise sources. We then move on to multi-qubit circuits, focusing on expectation-value circuits. We decompose the variance of the estimator into a sum of a statistical term and a bias floor. These are independently estimated with one additional run of the circuit. We test our method on a Variational Quantum Eigensolver for $H_2$  and show that we can predict the variance to within known error bounds. We go on to show that for IBM Pittsburgh's noise characteristics, at that instant, 7000 shots for the given circuit would have achieved a $\sigma^2 \approx 0.01$
\end{abstract}

\begin{keywords}
Quantum Computing, Quantum Circuit, Statistical Analysis
\end{keywords}

\maketitle

\section{Introduction}
Quantum computers are expected to provide exponential speed-ups in solving specific problems \cite{Shor_1997, Arrazola_2022}. However, the current era of quantum computing, often referred to as the Noisy Intermediate-Scale Quantum (NISQ) era, is characterised by devices with a limited number of qubits that are prone to errors and decoherence \cite{Preskill_2018}. Since the machines are noisy, we run each circuit multiple times to obtain a reliable result. The number of times we run the circuit is called the number of shots. The ideal number of shots is a trade-off between the precision of the result and the computational cost per shot.

In this work, we explore the problem of estimating the number of shots required to achieve the desired precision in the results, where the variance quantifies precision. There has been significant work on the convergence of VQEs with unbiased estimators across various algorithms~\cite{Peruzzo_2014, Sweke_2020}. These convergence studies, however, apply only to noiseless systems and we need to extended them to noisy systems for each run. There has also been work done on the distribution of noise in individual qubits~\cite{Burnett_2019, Carroll_2024}. Recently, there have been more empirical methods used, where the number of shots is determined by running the circuit multiple times and adjusting the number of shots until a satisfactory result is achieved through trial and error~\cite{Sung_2020, Brown_2020, Barron_2024}.

We aim to provide a procedure for estimating the number of shots required to achieve a desired variance in our results for NISQ-era machines, based on a statistical analysis of various noise sources and their impact on shot estimation. We rely on the Central Limit Theorem (CLT) as a tool to convert the distributions we obtain into a malleable form. We first apply the CLT to the noise sources of individual qubits and examine how it affects the variance in single-qubit experiments. Following this, since the general class of all multi-qubit quantum circuits would be intractable, we look at the subclass of circuits with expectation value type problems for Hermitian observables.

We work with four basic sources of noise that we assume are independent of each other. Hence, we treat them as four random variables $X_i$, such that $\text{Covar}(X_i, X_j) = 0,\,\forall\,i \neq j$. Additionally, since we assume independent sources of noise, it implies that our sources of noise are also independent of the circuit run. We deal primarily with four forms of well-studied noise sources: SPAM noise, amplitude damping ($T_1$), phase damping ($T_2$), and gate noise \cite{Geller_2021, Krantz_2019}.

While we focus on superconducting qubits, these methods may be generalised to other realisations of qubits, since they contain similar types of errors at different magnitudes \cite{An_2022, Strohm_2024}. Significant sources of error, such as correlated errors and crosstalk, are excluded from our analysis due to the lack of widely accepted models \cite{Bravyi_2018, Greenbaum_2017}.

The work is presented in two distinct parts.
\begin{itemize}
   \item First, we focus on single-qubit experiments to characterise individual noise sources and their impact on variance. We approach this from a 'bottom-up' physics-based model, slowly building up to a complete picture of variance from individual noise sources.
   \item Then we introduce a `top-down' empirical model to characterise variance in multi-qubit circuits, where the bottom-up approach becomes intractable. We demonstrate this method for a ground-state simulation of the $H_2$ molecule.
\end{itemize}

\section{Single Qubit Experiments}
We start with a single-qubit experiment of a Hadamard gate as the baseline and evaluate this circuit (Fig. \ref{circ:spam}) on a simulator and a QPU to obtain a coin toss distribution.
\begin{figure}
\begin{equation*}\scalebox{1.2}{
    \Qcircuit @C=1em @R=1em {
       \lstick{|0\rangle} & \gate{H} & \meter
    }
}\end{equation*}
\caption{Circuit baseline fair coin toss.}
\label{circ:spam}
\end{figure}
For a noiseless simulator, we expect a perfectly balanced distribution between $0$ and $1$ for the qubit. We find the expectation of the measurement $Z$ taken $n$ times as $\mu = E[X]=\sum _{i=1}^{n}x_{i}p_{i}$ such that $p_i$ is the probability of $x_i$, for all events $x_i$ in the sample space. We expect $\mu \rightarrow \mu_0$ as $n\rightarrow \infty$ with $\mu_0 = 0.5$ for a fair coin toss.

\newlength{\xfigwd}
\setlength{\xfigwd}{\textwidth}
\begin{figure}[h]
\includegraphics[width=1\linewidth]{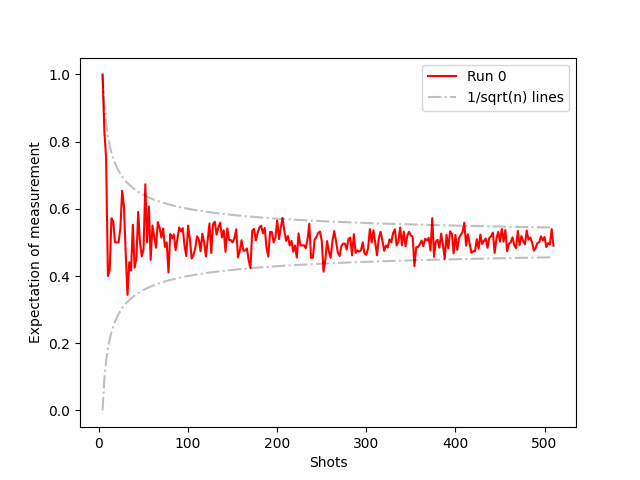}

\caption{As more shots are taken, the width of the distribution falls off as $1/\sqrt{n}$.}
\label{fig:baseline}
\end{figure}

Fig. \ref{fig:baseline} confirms our expectation of $\mu \rightarrow 0.5$ as $n$ increases. We also observe that the distribution's width falls off as $1/\sqrt{n}$ (overlaid for clarity). We will now use this plot as a baseline for all individual noise sources and see how they affect the distribution.

State Preparation And Measurement (SPAM) errors arise either during the preparation of the qubit or during a measurement on it. IBM's current noise simulators only allow us to simulate readout errors. For now, we \textbf{assume} they are negligible. Readout errors are made up of components: $p_{0 \rightarrow 1}$ and $p_{1 \rightarrow 0}$, the probability of a `0' being read as a `1' and the reverse, respectively. We write the modified expectation as,
\begin{align}\label{c:null}
\mu' = P(1) + p_{0 \rightarrow 1}P(0) - p_{1 \rightarrow 0}P(1).
\end{align}

This implies that when plotted with asymmetric readout error ($p_{0 \rightarrow 1} \ne p_{1 \rightarrow 0}$), we should see a shift in net expectation value. We have plotted the results for three cases:

\begin{itemize}
    \item Symmetric readout errors on IBM Qiskit noisy simulator: $(0, 0)$
    \item Symmetric readout errors on an independent noisy simulator presented in \cite{CERN_SIM}: $(0.33, 0.33)$
    \item Asymmetric readout errors on IBM Qiskit noisy simulator: $(0.33, 0.5)$
\end{itemize}

\begin{figure}[h!]
   \includegraphics[width=\linewidth]{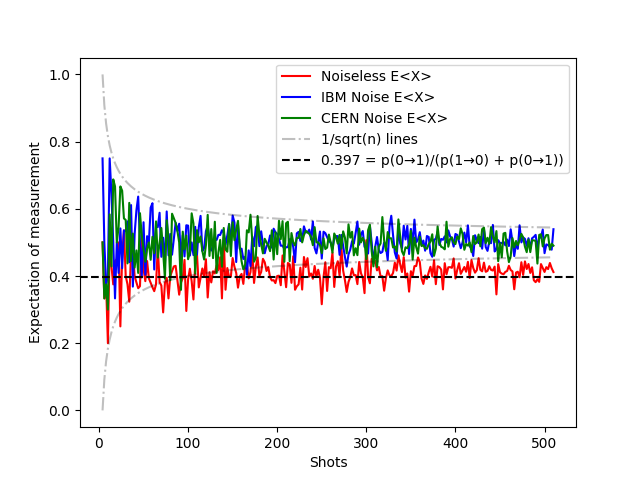}
   \centering

   \caption{When we add asymmetric noise to the coin toss, the distribution shifts downward. However, for symmetric noise, the mean is independent of the amount of noise.}
   \label{fig:spam}
\end{figure}

\noindent We see that the expectation for symmetric readout errors is the same as a fair coin. In contrast, in the asymmetric case, the expectation is shifted (see Fig. \ref{fig:spam}), as confirmed by our calculations.

\setlength{\xfigwd}{\textwidth}
\subsection{Central Limit Theorem and Relative Standard Deviation}
Having seen the standard deviation consistently fall off as $1/\sqrt{n}$, we now use a slightly modified metric of $\sigma/\mu$, or the relative standard deviation (RSD), to measure the spread of the distribution. RSD is dimensionless and can therefore be used to compare distributions across different scales. 

We first convert our distribution $X$ to a new distribution $\bar{X_w}$ by taking mean of windows of $w$ shots, such that $\bar{X_w} = \{\mu_{w, 1}, \mu_{w, 2}, \ldots, \mu_{w, n/w}\}$ as mean of subsets of values $[x_{i}, x_{i+w}] \in X$. From the classical central limit theorem, we know that $\bar{X_w}$ will be normally distributed as $w \rightarrow \infty$. We can then plot $\log(\text{RSD})$ vs $\log(w)$ for various values of $w$. The procedure as shown in Algorithm \ref{alg:cap}.

\begin{algorithm}[H]
\caption{Conversion to Normal Distribution.}
\label{alg:cap}

\begin{algorithmic}[1]
   \STATE $N \gets 256$ \COMMENT{Number of windows}
   \STATE $S \gets 2^{15}$ \COMMENT{Total shots}
   \STATE $W \gets \{2^2, 2^3, 2^4, 2^5, 2^6, 2^7\}$ \COMMENT{Window sizes}

   \FOR{each $w \in W$}
      \STATE Divide results into $N$ windows of size $w$
      \FOR{each window $i$}
         \STATE Compute mean $\mu_i$ for each window $i = 1, 2, \dots, N$
      \ENDFOR
      \STATE Compute mean $\mu$ and standard deviation $\sigma$ of $\{\mu_i\}$
      \STATE Store $\log(\frac{\sigma}{\mu})$ and $\log(w)$
   \ENDFOR
\end{algorithmic}
\end{algorithm}

\subsection{SPAM Noise}
\subsubsection{Simulation}
\begin{figure}[H]
   \centering
   \includegraphics[width=\linewidth]{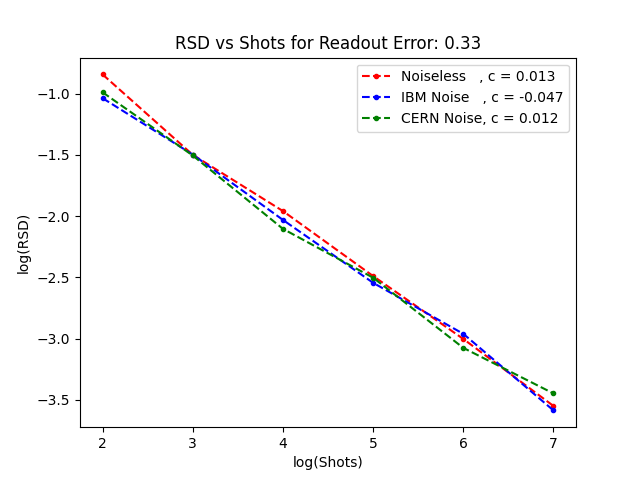}
    \caption{The standard deviation varies as $1/\sqrt{n}$ for each of the three cases simulated, noiseless, IBM noisy simulator and CERN noisy simulator.}
   \label{fig:clt_spam}
\end{figure}

As before, we run the SPAM experiments with three types of simulators: IBM's noisy simulator with no noise, IBM's noisy simulator with symmetric noise, and CERN's noisy simulator with symmetric noise. Here, symmetric noise is applied as $p_{0 \rightarrow 1} = p_{1 \rightarrow 0} = 0.33$. We can see the results of the experiments in Fig. \ref{fig:clt_spam}.

Treating the distribution as CLT, we now know we expect $c = y - mx$ where $m = -1/2$ and $c = \log(\sigma/\mu)$. We calculate the expected $c$ values for each of the three cases and compare them to the actual values. For a binomial distribution, we derive $c$ as:
\begin{align*}
  c &= y - mx \\
  &= \log(\sigma/\mu) + \frac{1}{2} \log(w) \\
  &= \log(\sqrt{wp(1-p)}/wp) + \frac{1}{2} \log(w) \\
  &= \frac{1}{2} \log\left(\frac{1 - p}{p}\right) = \frac{1}{2} \log\left(\frac{p_0}{p_1}\right). \numberthis \label{c:coin}
\end{align*}

Since $p \in (0,1) \implies c \in (-\infty, 0)$, and as $p_1 = p$ increases, the line shifts down.

\subsubsection{Hardware}
From Eq. \ref{c:coin} we expect
\begin{equation}
\Delta c = c_2 - c_1 = \frac{1}2 \log\left(
   \frac{1-p_1}{p_1} \frac{p_2}{1-p_2}
\right),
\end{equation}
\noindent for two different coin toss distributions with parameters $p_1, p_2$. We can treat $p_1=0.5$ as a fair coin, and $p_2$ as a biased coin, such that the probabilities of readout error have been accounted for in $p_2$ itself as done in (\ref{c:null}). $p_1, p_2$ now represent a fair coin and a SPAM noisy coin, respectively, letting us then write $c$ for SPAM noise as
\begin{align}\label{c:spam}
c_{\text{pred}} = \frac{1}2 \log\left(\frac
   {1 + p_{0 \rightarrow 1} - p_{1 \rightarrow 0}}
   {1 + p_{1 \rightarrow 0} - p_{0 \rightarrow 1}}
\right).
\end{align}

This implies that all symmetric SPAM noise models should have $c=0$ since $p_{0 \rightarrow 1} = p_{1 \rightarrow 0}$. Additionally, if the values of $p_{0 \rightarrow 1}, p_{1 \rightarrow 0} \rightarrow 0$ we should have $c \rightarrow 0$. We can also see that, if one of the readout errors is $0$ and the other is some $x\in (0,1)$, $c$ is highly sensitive to $x$ as $\frac{dc}{dx} = \pm 2/(x^2 - 1)$. This would imply that as we get closer to a higher probability of a 'uni-directional' readout error, our variance explodes. We would therefore prefer both small and symmetric readout errors.

We test our result for $c_{\text{pred}}$ on the IBM Torino machine and see if we can predict the $c$ values for the qubits. From Table~\ref{table:spam_table}, we can see that we can predict the $c$ values within $0.01$ of the actual values from just the calibration data. We can also see that the $c$ values are generally also more negative since we expect more noise than we account for with only SPAM. It is unknown why Qubit 61 shows a higher error in our $ c$-value prediction than the other qubits. We hypothesise that Qubit 61 either had additional sources of noise or that the calibration data was outdated, given the 2 days of queue time required to account for our error.
\begin{table}[htb]
\begin{center}
\caption{Qubit parameters from IBM Torino's calibration and comparable values of $c$.}\label{table:spam_table}
\begin{tabular}{|c|c|c|c|c|c|}
   \hline
   Qubit & $T_1$ & $p_{0 \rightarrow 1}$ & $p_{1 \rightarrow 0}$ & Expected $c$ & Actual $c$\\
   \hline
   61 & 232 & 0.0099 & 0.006 & -0.01 & -0.13 \\
   129 & 232 & 0.0186 & 0.014 & -0.012 & -0.012 \\
   5 & 232 & 0.0350 & 0.0428  & 0.023 & 0.033 \\
   \hline
\end{tabular}
\end{center}
\end{table}
\subsection{$T_1$ noise}
$T_1$ decay increases the probability of a qubit going from $|1\rangle$ to $|0\rangle$ over time. We can test for $T_1$ by modifying our circuit to wait for a known time before measurement, as shown in Fig~\ref{eq:t1_circ}, then adding a decay to our coin toss experiment. We will apply `wait' at time multiples of $\delta t$ (the time unit for pulse operations, which IBM refers to as `$\text{dt}$') for single-qubit gates on the IBM Torino system. Our `wait' gate will be implemented as an identity ($\text{Id}$) gate, therefore $\text{wait}(n) = n \cdot \delta t$. An initial $H$ gate, along with $n$ wait gates give us a total circuit runtime of $t = (n+1)\delta t.$

\begin{figure}
\begin{equation*}
\Qcircuit @C=1em @R=1em {
   \lstick{|0\rangle} & \gate{H} & \gate{\text{wait}(t)} & \meter
}
\end{equation*}
\caption{Circuit for $T_1$ noise with variable wait.}
\label{eq:t1_circ}
\end{figure}

Let us define $\epsilon = e^{-t/T_1}$ \cite{Krantz_2019}, where $t$ is the runtime of the circuit. We write modified expressions
\begin{align}
  p_0' = p_0 + p_1 \epsilon, \\
  p_1' = p_1 - p_1 \epsilon,
\end{align}

\noindent and substitute them back into the SPAM formula, giving us,
\begin{align}\label{c:t1}
c_{\text{pred}} = \frac{1}{2} \log\left(\frac{
   b (1 - p_{0 \rightarrow 1}) + p_{1 \rightarrow 0} }{
   b (p_{0 \rightarrow 1}) + 1 - p_{1 \rightarrow 0}
}\right), \text{ where } b = \frac{1 + \epsilon}{1 - \epsilon}.
\end{align}

\noindent We observe that when $\epsilon \rightarrow 0 \implies b \rightarrow 1$ and we regress to the SPAM noise formula.\

\begin{figure}[htb]
   \centering
   \includegraphics[width=\linewidth]{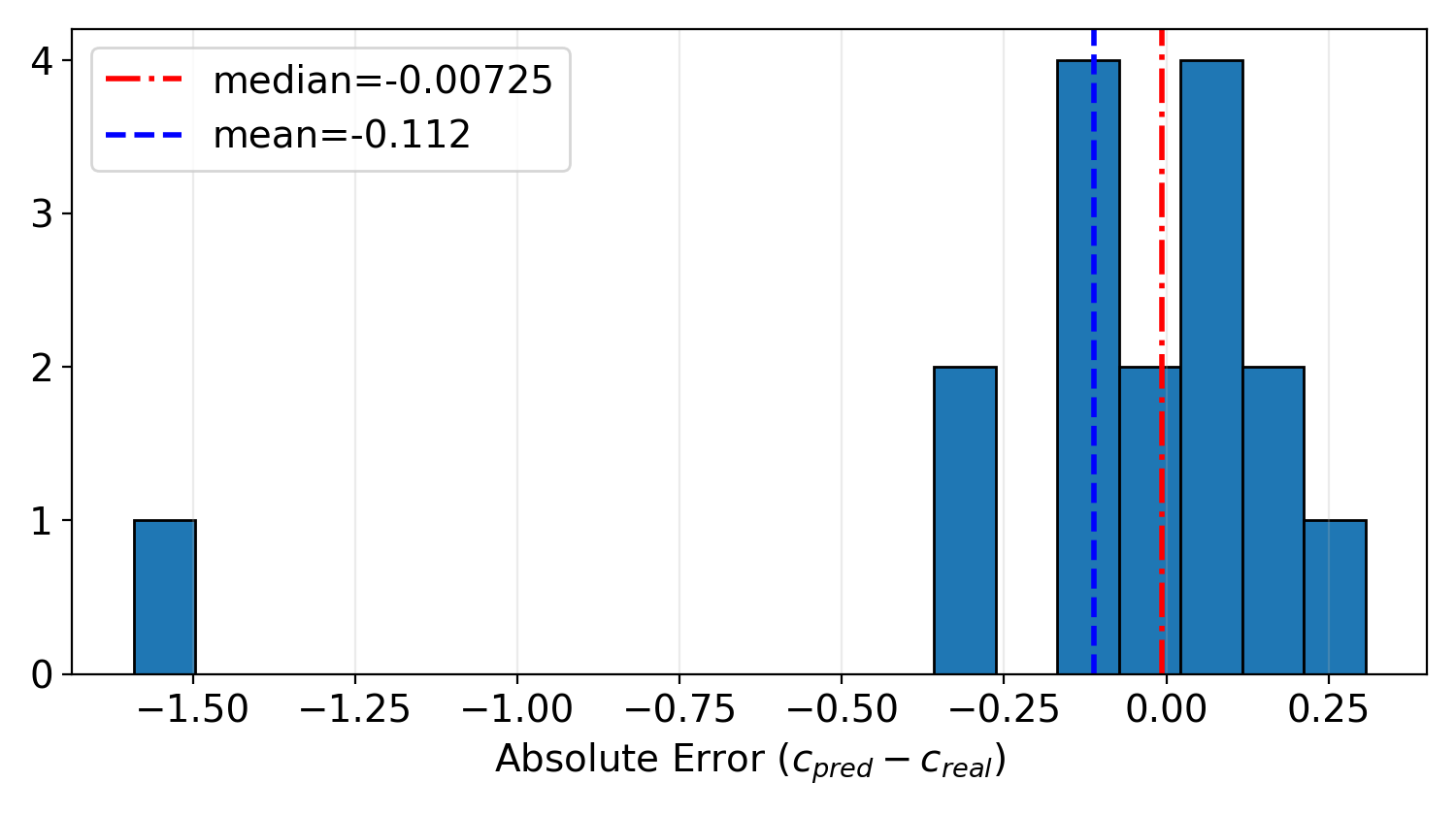}

   \caption{As we can see from the histogram, we have a median error very close to 0 (-0.007), and a mean error of 0.1 despite an outlier event. If we were to remove the outlier event, the mean would be -0.013.}
   \label{fig:t1_map}
\end{figure}

We compare $c_{\text{pred}}$ to $c_{\text{real}}$ as $\Delta_\epsilon c = |c_{\text{pred}} - c_{\text{real}}|$ after running the modified coin toss circuit in Fig. \ref{eq:t1_circ} on IBM Torino's hardware. A histogram of the results is shown in Fig~\ref{fig:t1_map}.
We can conclude that we can predict the shift in the intercept due to $T_1$-induced errors to reasonable accuracy and can apply the same to $T_2$ noise.

\subsection{$T_2$ noise}
$T_2$ decay makes the probability of a qubit going from any known phase, say $|+\rangle$, to a uniform superposition of $|+\rangle$ and $|-\rangle$ increase over time. Equivalently, the probability of a $Z$ gate being applied to a qubit increases as $p \propto 1 - e^{-t/T_2}$. As before let us define $\epsilon_1 = e^{-t/T_1}, \epsilon_2 = e^{-t/T_2}$. If we then apply a $T_1, T_2$ noisy channel onto a density matrix, we get the resultant state as \cite{Krantz_2019}
\begin{equation}
\rho_{T_1 + T_2} = \begin{pmatrix}
1 + (|a|^2 - 1)\epsilon_1 & ab^* \epsilon_2\\
a^*b \epsilon_2 & |b|^2 \epsilon_1
\end{pmatrix}.
\end{equation}

\noindent Multiplying by the $H$ gate matrix, we get probabilities for the coin-toss experiment with $T_2$ in the $H$ basis.
\begin{align}\label{c:t2}
   c =  \frac{1}{2} \log\left(\frac{
      b (1 - p_{0 \rightarrow 1}) + p_{1 \rightarrow 0} }{
      b (p_{0 \rightarrow 1}) + 1 - p_{1 \rightarrow 0}
   }\right), \text{ where }
   b = \frac{1 + \epsilon_2}{1 - \epsilon_2}.
\end{align}

\begin{figure}[htb]
   \centering
   \includegraphics[width=\linewidth]{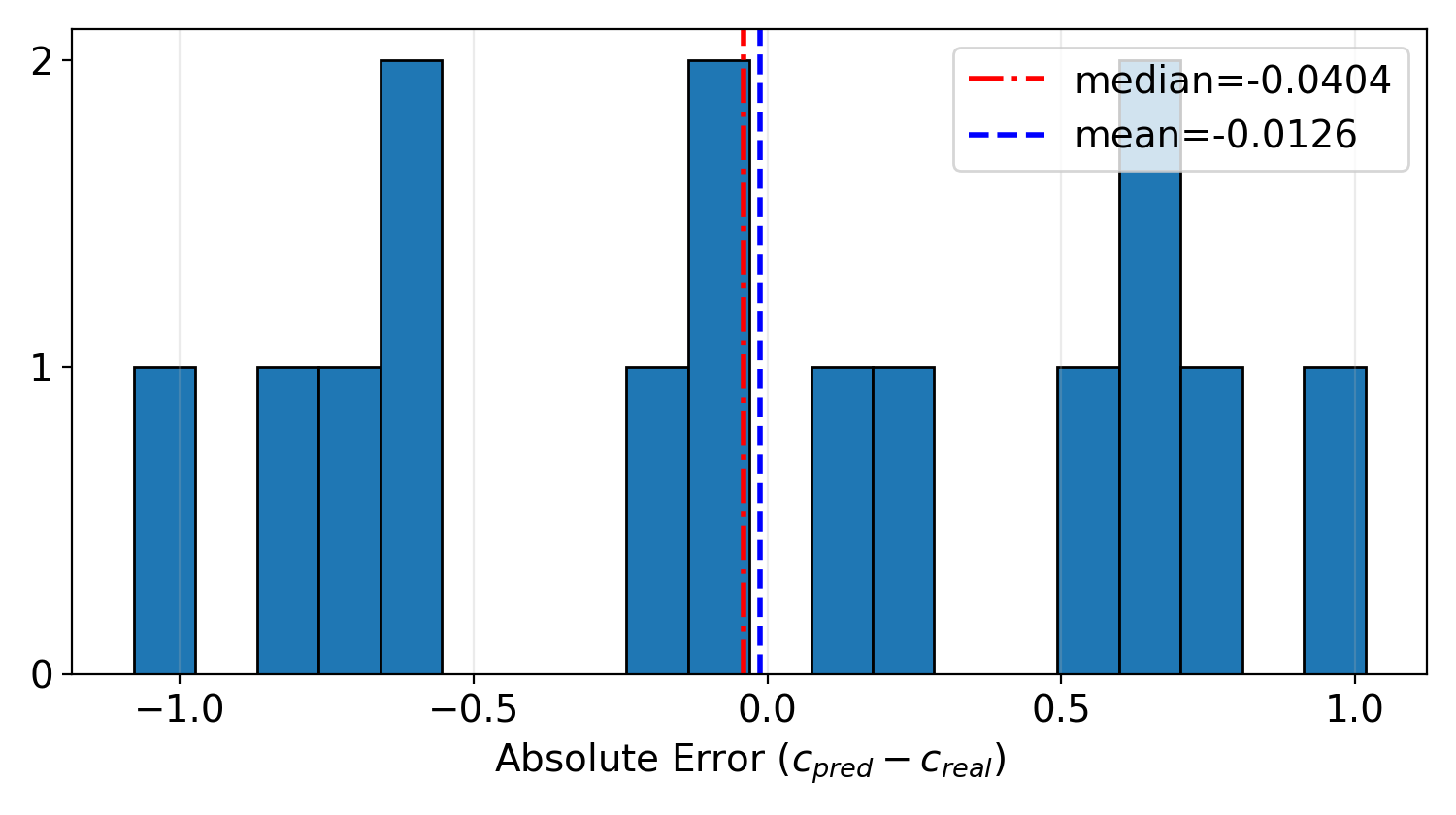}

   \caption{We can see that we are able to predict $T_2$ noise with the mean and median of errors within $\approx1\%$ despite a higher variance.}
   \label{fig:t2_map}
\end{figure}

\begin{figure}
\begin{equation}\scalebox{1.2}{
\Qcircuit @C=1em @R=1em {
   \lstick{|0\rangle} & \gate{H} & \gate{\text{wait(t)}} & \gate{H} & \meter
}
}\end{equation}
\caption{Circuit for $T_2$ noise.}
\label{eq:t2_circ}
\end{figure}

We can see that, when predicting the intercept due to $T_2$ noise, the noise has a higher spread but remains centred around $0$, Fig~\ref{fig:t2_map} is the output of running the circuit in Fig~\ref{eq:t2_circ}. This could be attributed either to the additional operations required to change bases or to other unaccounted-for sources.

By virtue of being constructed from the Central Limit Theorem, one may even generate confidence levels for their obtained intervals with little effort. We may choose to adopt an even broader range of acceptable $\Delta_\epsilon c$ by accounting for uncertainty in $ T_1$ and $ T_2$ themselves as follows. Let's say the uncertainty in $T_2$ is some $\sigma_{T_2}$, then the uncertainty in $c$ is given by,

\begin{equation}
\sigma_c = \frac{\partial c}{\partial T_2} \sigma_{T_2}.
\end{equation}

We have defined $c$ for the $T_2$ experiment as (\ref{c:t2}). Let us say we were to ignore the readout error for now and account only for $T_2$, we can write,
\begin{equation}
\sigma_c = \frac{\partial c}{\partial T_2} \sigma_{T_2}= \frac{t e^{t/T_2}}{T^2 (e^{2t/T_2} -1)} \sigma_{T_2}.
\end{equation}
\noindent Therefore if we have $T_2 = 500\,\delta t$ at some circuit depth $t = 100\,\delta t$, then for $\sigma_{T_2} = 20\,\delta t$
we will have $\sigma_c = 0.02$. We can see that as $T_2$ increases, the uncertainty in $c$ falls as $ 1/T_2^2$. Therefore, for sufficiently large $T_2$, the uncertainty in $c$ will be very small. We can also see that $t \rightarrow 0$ implies $\sigma_c \rightarrow 0$.

\subsection{Gate Noise}
\begin{figure}[htb]
   \centering
   \includegraphics[width=\linewidth]{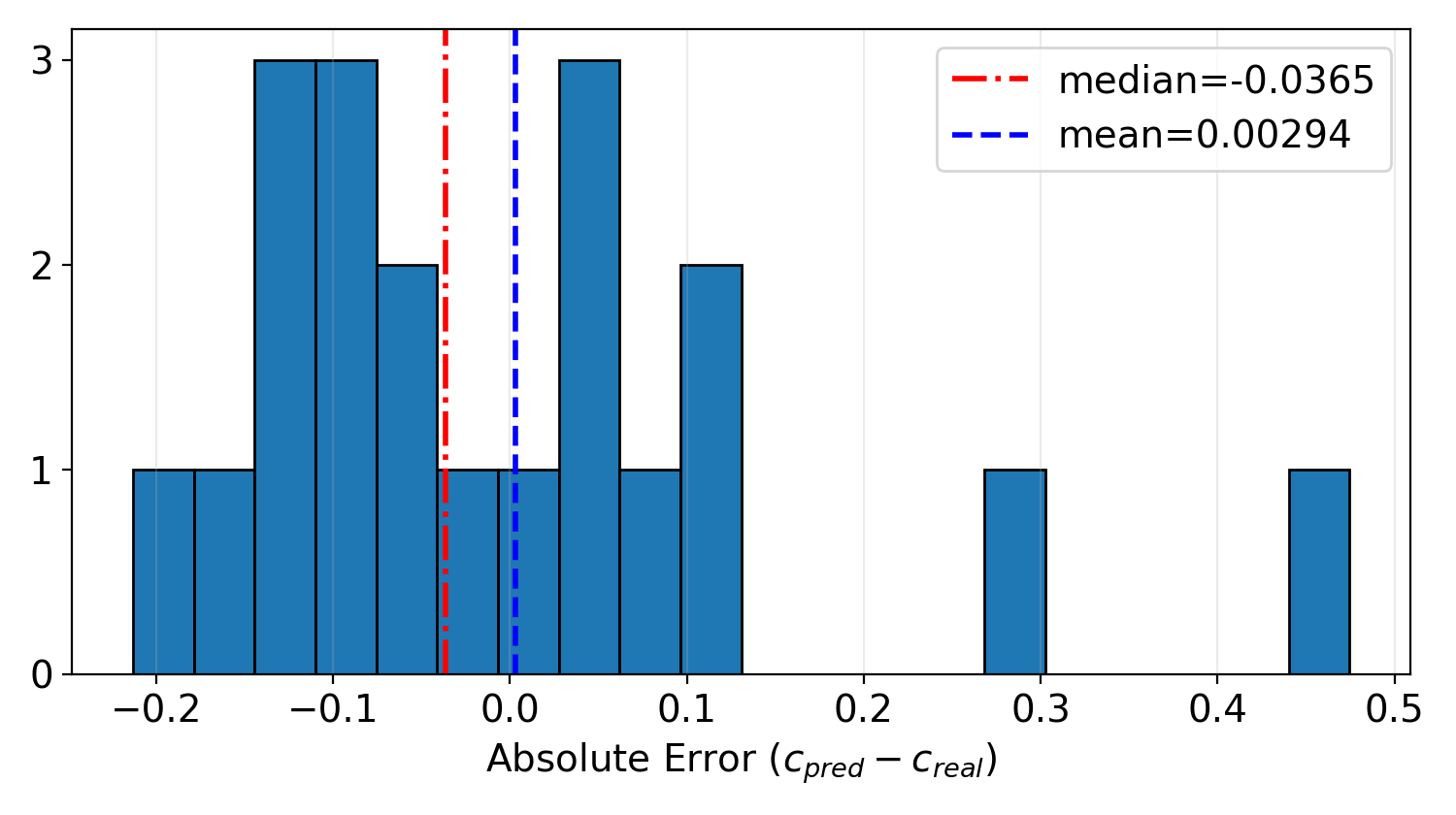}
   \caption{There seems to be a slight bias towards under-prediction; however, our mean and median error both remain ~0.1. It is worth noting that these errors are after we have dropped accounting for SPAM noise.}
   \label{fig:ge_pred}
\end{figure}

Assuming no correlated errors, we can model gate errors as depolarising noise channels, since uncorrelated errors are Pauli errors. As before, we can start with the Jaynes-Cummings model of a qubit interacting with an environment, then apply the depolarising channel. In interest of simplification, we apply $\alpha = \beta = 1/\sqrt{2}$ to get the density matrix as:
\begin{equation}
\begin{bmatrix}
\epsilon_1(1+p_z) + 1 - \tfrac{1}{2} \epsilon_1
& - (p_y + p_z) \epsilon_2 + \tfrac{1}{2} \epsilon_2 \\
- (p_y + p_z) \epsilon_2 + \tfrac{1}{2} \epsilon_2
& (1-p_z) (1 - \epsilon_1) + \tfrac{1}{2} \epsilon_1
\end{bmatrix}
\end{equation}
\noindent where $p_x, p_y, p_z, 1-p_x- p_y- p_z$ are the probabilities of $X, Y, Z, I$ errors, respectively. We can then estimate the probabilities for the coin toss experiment from the matrix. What remains is to obtain $p_x, p_y, p_z$ from the gate error rates in the calibration data. It is very difficult to obtain exact values for each error independently; therefore, we will assume that each error is equally likely, such that $p_x = p_y = p_z = p$.
From randomised benchmarking\cite{EPLG2023}, we can obtain the value for Error Per Layered Gate (EPLG, $E$), which recovers its value as $E = p_x + p_y + p_z = 3p$, derived in Appendix \ref{Appendix:eplg}. Therefore, we can write $p = E/3$. For successive errors along a given axis, we can multiply the probabilities of no error along that axis. Therefore, for $k$ gates, and assuming $p \ll 1$ or large $k$, we can write the effective probabilities as
\begin{align*}
P_k &= \left(1 - (1 - p)^{k}\right), \\
&\approx kP. \numberthis
\end{align*}
It is known that the error of the binomial approximation $(1 - x)^n \approx 1 - nx$ is bounded by $\epsilon_{\text{Binomial}} \leq \frac{n(n-1)}{2}x^2$ for $x \in [0,1], n \ge 2$. If for a large $k$, we were to set our error boundary as $\epsilon_{\text{Binomial}} < 0.5$, we would obtain $k \leq \frac1p$. It is worth noting that, outside this boundary, the circuit may still be valid and have small error rates, but the approximation error may be significant enough to generate incorrect predictions.
We can now substitute in $p_x = p_y = p_z = kE/3$ into the density matrix and obtain probabilities for the coin toss experiment. Substituting these probabilities into the SPAM formula, we get:
\begin{align}
\mathbb{P}(1) = \frac{\left(4Ek e^{\tfrac{t}{T_1}} - 4Ek + 3\right)e^{-\tfrac{t}{T_1}}}{6}.
\end{align}
With this value of $\mathbb{P}(1)$ we can obtain $c_{\text{pred}} = 0.5 \log(\mathbb{P}(0)/\mathbb{P}(1))$. When run on IBM Torino's hardware, we obtain the histogram of errors as in Fig. \ref{fig:ge_pred}.
With individual noise sources characterised, we now move on to multi-qubit circuits. For general circuits, it is intractable to derive variance from individual qubit variances. Hence, we focus on expectation-value circuits, where we can directly measure the observable's variance.

\section{Circuit Experiments}
Consider $k$ independent bits $X_i$ where:
\begin{itemize}
    \item $X_i = 1$ transitions to $0$ with probability $p$
    \item $X_i = 0$ transitions to $1$ with probability $q$
\end{itemize}
The variance of each bit $X_i$ is given by
\begin{align}
\Var{X_i} = (1 - p + q)(p - q) \cdot 2^i.
\end{align}
If the bits are combined bitwise into an integer (e.g., $101010_2 = 42_{10}$), the variance of the resulting number scales exponentially, irrespective of $p$ and $q$. However, if they are interpreted as fractional bits as in the phase estimation algorithm (e.g., $0.01101011_2 \approx 0.42_{10}$), where each bit represents a fraction of increasing precision, the resulting variance remains small.
Further, if we construct a composite random variable $Y = X_1 + X_2$, its variance is given by
\begin{align}
\Var{Y} = \Var{X_1} + \Var{X_2} + 2\,\text{Cov}(X_1, X_2).
\end{align}
Note that, we cannot ignore the covariance term here, as the bits may be entangled. Finally, if we define a random variable $Z = X_1 \otimes X_2$, with a variance
\begin{align}
\Var{Z} = (\Var{X_1} + \mu_{X_1}^2)(\Var{X_2} + \mu_{X_2}^2) - \mu_{X_1}^2 \mu_{X_2}^2.
\end{align}
Depending on how the bits are combined, the variance of the resulting variable can differ significantly. Therefore, in the general case, we cannot derive the variance of a circuit solely from the variances of its constituent bits. Even if we knew the precise mechanism of combination or had a universal method, we would still require the covariance between every pair of qubits, which is a combinatorial explosion in the number of qubits.
\subsection{Expectation Value Circuits}
We study circuits that measure expectation values of observables, to which many problems can be mapped. For an observable $O$ measured on a state $|\psi\rangle$ prepared by an ansatz,
\begin{equation}
\sigma^2 = \Var{O} = \langle O^2 \rangle - \langle O \rangle^2,
\end{equation}
while the variance of the estimator $\bar{E}_N$ over $N$ samples is
\begin{equation}
\Var{\bar{E}_N} = \frac{\Var{O}}{N} = \frac{\sigma^2}{N}.
\end{equation}
We adopt a simple, physically motivated decomposition of the estimator variance into statistical variance plus a bias floor:
\begin{align}\label{eq:anmplusb}
\Var{\bar{E}_N} = \frac{A}{N} + B,\quad A \approx \sigma^2,\; B \ge 0,\;
\end{align}
\noindent where $A$ captures the intrinsic quantum variance scaled by finite sampling, and $B$ captures device- and circuit-dependent bias (systematic error floor).
\subsection{Application}
Additionally, if $N$ shots of one circuit are one run, then only two physical runs are needed to estimate variance at all shot counts by reusing the same data. One, for $\langle O \rangle$ and another for $\langle O^2 \rangle$. 
\subsection{$H_2$ VQE}\label{app:1}
\begin{figure}[htb]
   \centering
   \includegraphics[width=\linewidth]{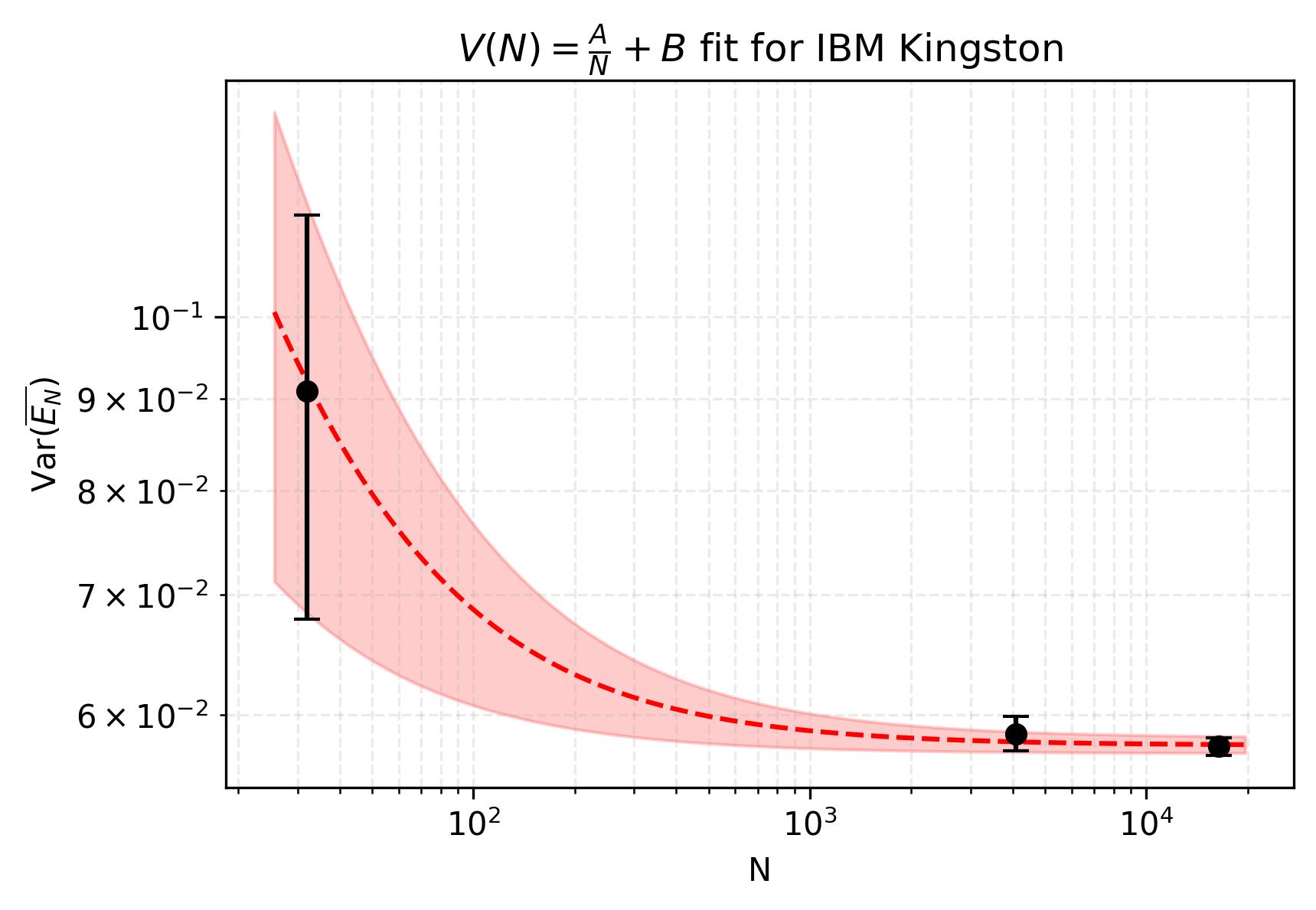}
   \caption{IBM Kingston: three-point fit to $\Var{\bar{E}_N}=A/N+B$. Error bars show $\mathrm{SE}(s^2)$. The predicted variance at other $N$ is consistent with observations within uncertainty.}
   \label{fig:fit_kingston}
\end{figure}

\begin{figure}[htb]
   \centering
   \includegraphics[width=\linewidth]{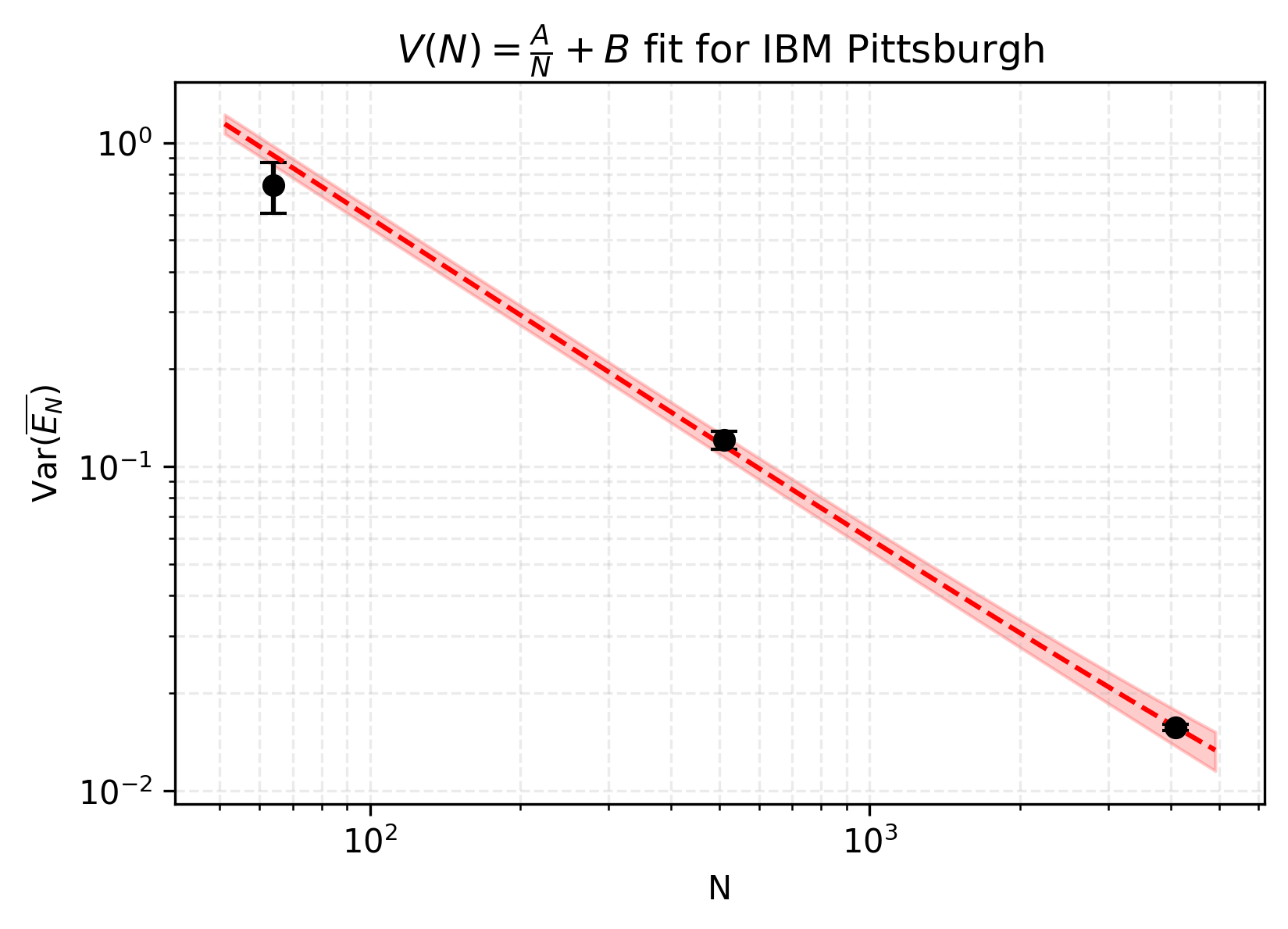}
   \caption{IBM Pittsburgh: three-point fit to $\Var{\bar{E}_N}=A/N+B$. Error bars show $\mathrm{SE}(s^2)$. Predictions align with measured variances.}
   \label{fig:fit_pittsburgh}
\end{figure}

We construct a standard $H_2$ VQE with a 4-qubit Hartree–Fock circuit in the STO-3G basis \cite{Sun_2020}. The observable and the excitation-preserving circuit are shown in Appendix \ref{appen:two}. Since we estimate energy from $\pm 1$-eigenvalue measurements of the observable, we measure $\langle O \rangle$ and $\langle O^2 \rangle$ to compute the variance at each shot count.
To quantify uncertainty in the estimated variance, we also compute its standard error (SE). Since $\bar{E}_N$ is approximately normal, the sample variance $s^2$ follows a scaled $\chi^2$ distribution with $N-1$ degrees of freedom. Thus, the standard error of the variance is
\begin{align}\label{eq:SE}
\mathrm{SE}(s^2) &= s^2 \sqrt{\frac{2}{N-1}}, \quad \\ 
\text{where }\, s^2 &= \frac{1}{N-1} \sum_{i=1}^{N} (E_i - \bar{E}_N)^2.
\end{align}
This allows us to place confidence bounds on $\Var{\bar{E}_N}$ and its fitted form in Eq.~\eqref{eq:anmplusb} with no additional measurements.
\begin{figure}[htb]
   \centering
   \includegraphics[width=\linewidth]{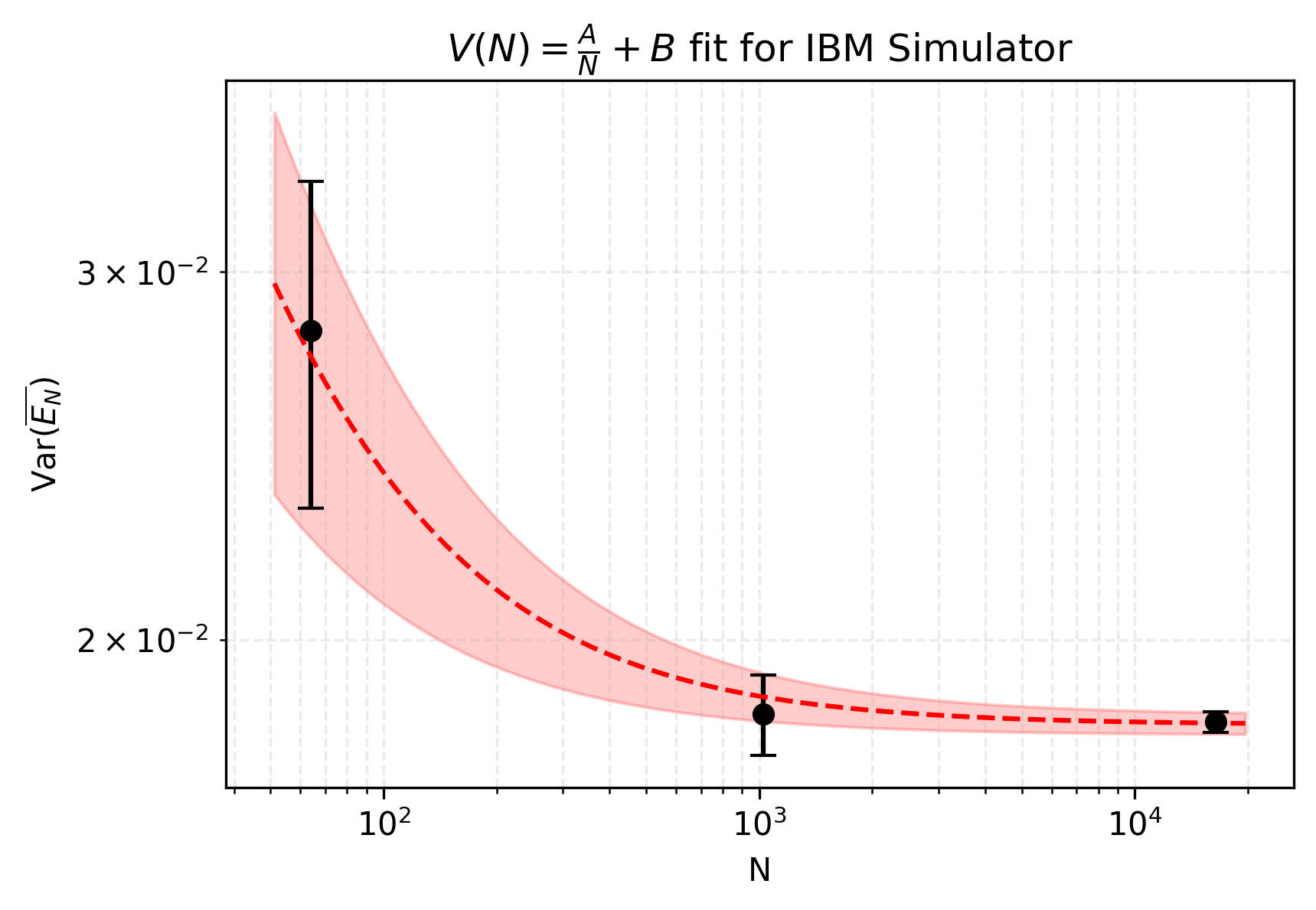}
   \caption{Three-point fit of $\Var{\bar{E}_N}$ to $A/N + B$ on a simulator. Error bars denote $\mathrm{SE}(s^2)$. The fitted curve predicts variance at unseen shot counts within uncertainty.}
   \label{fig:h2_sim}
\end{figure}
We then repeat the same three-point procedure on IBM Kingston (Fig. \ref{fig:fit_kingston}) and IBM Pittsburgh (Fig. \ref{fig:fit_pittsburgh}).
Note that variance is sensitive to device conditions (qubits chosen, layout, time of execution). While $A$ and $B$ provide an operational characterisation of a given run, their values may differ across devices and over time. For similar circuits on comparable hardware, $A$ may vary with state preparation while $B$ often reflects the device's systematic floor.
The advantage of a bias-plus-variance model is that it also provides us with the minimum achievable variance $B$ as $N \rightarrow \infty$. This allows us to determine whether increasing shots is worthwhile. Once the equation is regressed, we can also solve the inverse problem: if our desired final variance is $\sigma^2$, we can solve for $N = \frac{A}{\sigma^2 - B}$. For the $H_2$ experiment at $\sigma^2=0.01$, on the IBM Pittsburgh hardware (Fig. \ref{fig:fit_pittsburgh}), this would imply ~7000 shots.
From Fig. \ref{fig:fit_kingston}, we see that the first point, with a very low shot count ($N=32$), has a very high SE. Since it is known that the $\text{SE}(\sigma^2) \propto \sigma^2$ by the scaling factor of $\sqrt{2/(N-1)}$, should one wish to enforce a maximum SE of say, $p\%$, one can easily rearrange Eq.\eqref{eq:SE} to get
\begin{equation}
   N > \frac{2\cdot 10^4}{p^2}.
\end{equation}

\section{Conclusion}
We have presented a framework for understanding and predicting the shot-variance relationship in noisy quantum computers, with a focus on providing practical tools for NISQ users. We first developed a bottom-up analytical model to predict this systematic error floor (parameter $c$) from first-principles calibration data (SPAM, $T_1$, $T_2$, gate errors). While effective for single-qubit circuits, we confirmed that the complex covariance terms in multi-qubit systems make this predictive approach intractable, a known challenge in the field. To circumvent this intractability, we move to a top-down, empirical methodology for complex, multi-qubit circuits.
We proposed that the total estimation error (Mean Squared Error) follows the physically-motivated model: $MSE = \frac{\sigma^2}{N} + B^2$. Here, $\sigma^2$ is the intrinsic statistical variance of the observable, and $B^2$ is the squared systematic bias (the "error floor") that is the practical equivalent of our $c$ parameter. By fitting experimental data from VQE case studies ($H_2$ on IBM's Kingston and Pittsburgh devices) to this model, we can empirically extract both parameters, $\sigma^2$ and $B^2$. With both parameters known, we have a definite relationship between shot count $N$ and error. This allows NISQ users to optimally allocate their finite shot budget to meet desired accuracy targets.
The open problems that remain include developing a deeper theoretical understanding of the causes of the systematic bias floor $B^2$ in multi-qubit circuits and identifying the fundamental parameters that govern its behaviour.

\section{Appendix}
\subsection{Hydrogen VQE}\label{appen:two}
The approximate Hamiltonian for $H_2$ in STO-3G basis is\cite{McArdle_2020}
\begin{align*}
   H = 0.045 YYYY + 0.045 XXYY + 0.045 YYXX \\
   + 0.045 XXXX + 0.120 IIZZ + 0.120 ZZII \\
   + 0.166 ZIIZ + 0.166 IZZI + 0.168 IZIZ \\
   + 0.170 IIIZ + 0.170 IZII + 0.17 ZIZI \\
   - 0.219 ZIII - 0.219 IIZI - 0.815 IIII. \numberthis
\end{align*}

\begin{figure}[htb]
   \includegraphics[width=\linewidth]{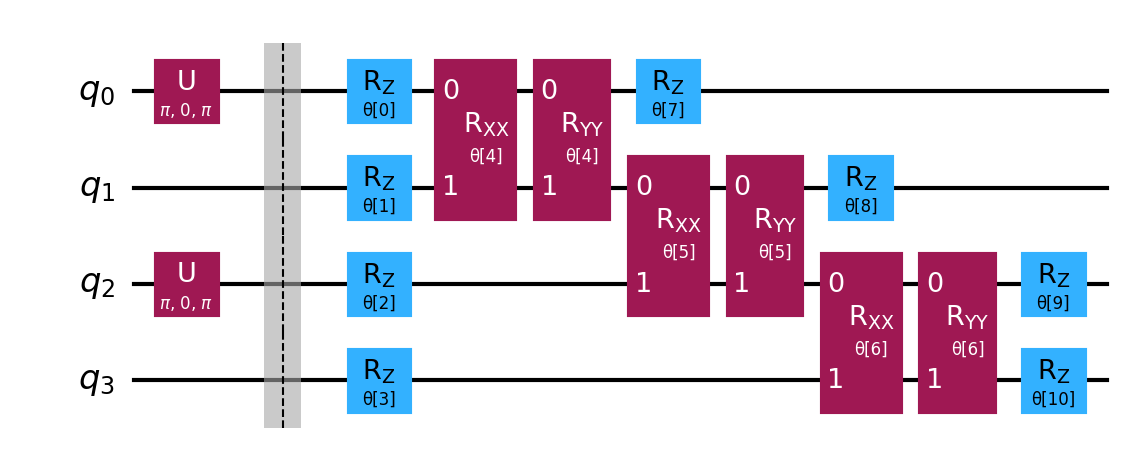}.

   \caption{$\text{H}_2$ Hartree Fock initial circuit, followed by a hardware-efficient excitation-preserving ansatz.}
   \label{fig:h2vqe}
\end{figure}

\noindent The circuit generated for the same is as shown in Fig. \ref{fig:h2vqe}.

\subsection{EPLG Derivation}\label{Appendix:eplg}
Given probabilities $(p_X,p_Y,p_Z, 1-p_X-p_Y-p_Z)$ for $X,Y,Z,I$ errors respectively, for a state $\rho$, the depolarising channel is defined as
$$
\Lambda(\rho)=\rho + p_X X\rho X + p_Y Y\rho Y + p_Z Z\rho Z.
$$
The entanglement fidelity $F_e(\mathcal{E})$ for a channel $\mathcal{E}$ is defined as\cite{Nielsen2010}
\begin{align}
F_e(\mathcal{E}) &= \frac{1}{N^2}\sum_i|\mathrm{tr}K_i|^2\\
\implies F_e(\Lambda) &= 1-(p_X+p_Y+p_Z).
\end{align}
Further, Error Per Layered Gate (EPLG/E) is defined as\cite{EPLG2023}
\begin{align}
E=1 - \text{LF}^{\frac1n} \text{ where } \text{LF} = \Pi_i^n F_i,
\end{align}
\noindent where LF is the Layered Fidelity over $n$ gates. We may substitute $F_e(\Lambda_i)$ for each gate to get a new LF as
\begin{align}
\text{LF} = \Pi_i (1 - (p_{X_i} + p_{Y_i} + p_{Z_i})).
\end{align}
At this point, as a crude approximation, we may reasonably pick $p_{X_i} = p_{Y_i} = p_{Z_i} = p$ for all $i$. Therefore,
\begin{align}
\text{LF} &= (1 - 3p)^n, \\
\implies E &= 1 - (1 - 3p) = 3p.
\end{align}

\bibliographystyle{unsrt}
\bibliography{ref}

\EOD
\end{document}